\documentclass[12pt,thmsa]{article}

\usepackage{amsfonts}
\usepackage{graphicx}
\usepackage{epsfig}
\usepackage{graphics}
\makeatletter
\newcommand{\row}[1]{\mathord{\buildrel{\lower3pt\hbox{$\scriptscriptstyle\rightarrow$}}\over #1}}

\newcommand{\dyadic}[1]{\mathord{\dyadic@rrow{#1}}}
\newcommand{\dyadic@rrow}[1]{
\begin{picture}(12,12)(-1,0)
\put(-2,12){\makebox(0,0)[t]{$\scriptscriptstyle\downarrow$}}
\put(-2,12){\makebox(0,0)[l]{$\scriptscriptstyle\longrightarrow$}}
\put(5,0){\makebox(0,0)[b]{$#1$}}
\end{picture}
}
\newcommand{\bra}[1]{\bigl\langle #1 \bigr|}
\newcommand{\ket}[1]{\bigl| #1 \bigr\rangle}

\begin{document}
\begin{center}
{\Large Information transfer and orthogonality speed via
pulsed-driven qubit}\\
\vspace{0.5cm}
N. Metwally and S. S.Hassan\\
Department of Mathematics  , College of Science, University of Bahrain,\\
P. O. Box 32038 Kingdom of Bahrain \\
 Nmetwally@gmail.com, Shoukryhassan@hotmail.com\\

\end{center}

\begin{abstract}
We investigate the transfer and exchange information between a
single qubit system excited by a rectangular pulse. The dynamics
of the system is treated  within and outside  rotating wave
approximation (RWA). The initial state of the qubit plays an
important role for sending information with high fidelity. Within
RWA, and as the fidelity of the transformed information increases
the exchange information with the environment increases. For
increasing values of atomic detuning, the fidelity decreases
faster and the exchange information has an upper limit. Outside
RWA,the fidelity of the transformed information increases as one
increases the perturbation parameter. However the exchange
information is very high compared with that within RWA. The
orthogonality speed of the travelling qubit is investigated for
different classes of initial  atomic state settings and  field
parameters.
\end{abstract}

\section{Introduction}

Within  the context of quantum information theory,quantum objects
are considered as carriers  of information \cite{Audr}. A
two-state system, such as single 2-level atom, called single qubit
is one basic example of quantum information unit. Qubits represent
a fundamental aspect of quantum computer \cite{Audr,Laf}.
Dynamical properties of single qubits have been investigated in
different directions. Very recently, single control qubit Shor
algorithm for the case of static imperfections induced by residual
couplings between qubits was formulated \cite{Ign}.

Cryptographic applications of single-qubit rotations in quantum
public-key cryptography  has been discussed in references
\cite{Geo}-\cite{Chr}. The use of   single photon qubit as a
quantum encoder for single-photon qubits was reported in
\cite{Pitt}. The authors in \cite{Metwally} have used the charged
qubit pair to perform quantum teleportation. The speed of
communication using single qubit has been treated in classical and
quantum framework \cite{Mar,Borrs,Metwally2}.

 Dynamical and spectral
properties of a short pulsed -driven qubit in the absence of
dissipation processes has been studied  by  many
researchers(\cite{Shore}-\cite{Jos} and refs therein). In
\cite{shukry}, the authors have  investigated the fluorescence
spectrum of a rectangular pulsed-driven single  qubit within and
outside  the rotating wave approximation (RWA, where fast
oscillatory terms are dropped )  for different initial atomic
states. Specifically, for  initial atomic coherent state, the
transient fluorescence  spectrum exhibits asymmetric Rabi
splitting which turns to "ringing" for large pulse area. The
ringing behavior is attributed to the initial finite coherent
dispersion and interference process between amplitude spectra of
the atomic inversion and polarization variables. Properties of
single qubit rotation operations  using simple RF pulses have been
investigated in \cite{Mor}. Also, measure of the errors in single
qubit   rotations by pulsed electron paramagnetic resonance has
been reported in \cite{Mor1}.

In the present work, we investigate the dynamics  of the coded
information in a single qubit subject to a rectangular pulse. We
quantify the exchange information between the qubit and the
environment(i.e the pulse). Also, we quantify the speed of quantum
communation by evaluating  the speed of orthogonality of the
density operator.

The paper is organized as follows. In Sec.2, we present the model
and its  exact operator  solution  within  the RWA and its
iterative solution outside RWA \cite{shukry}. The fidelity and the
exchange information is the subject of Sec.3. The speed of
orthogonality  and hence  the speed of transfer information is
discussed in Sec.4. Finally, a summary is given in Sec.5.

\section{The Model and its solutions}
We consider  a single 2-level atom (qubit) of transition frequency
$\omega_a$ interacts with a rectangular pulse of circular
frequency $\omega_l$ in the absence of any atomic dissipation. The
full Hamiltonian of this system outside the RWA (in units of
$\hbar=1$) is given by \cite{shukry},
\begin{equation}\label{Ham}
\hat{H}=\omega_a\hat{S_z}+\frac{\Omega(t)}{2}\Bigl\{(\hat{S_{+}}+\hat{S_{-}})(e^{-i\omega_l
t}+e^{i\omega_l t})\Bigr\},
\end{equation}
where  the spin-$\frac{1}{2}$ operators $\hat{S_{\pm}},\hat{S_z}$
obey the $Su(2)$ algebra,
\begin{equation}\label{Com}
[\hat{S_{+}}, \hat{S_{-}}]=2\hat{S_{z}},\quad
[\hat{S_{z}},\hat{S_{\pm}}]=\pm\hat{S_{\pm}}.
\end{equation}
For a  rectangular laser  pulse of duration $T$,  the Rabi
frequency  $\Omega(t)=\Omega f(t)$ with $f(t)=1$ through the
interval of time $t\in[0,T]$ and $f(t)=0$ otherwise, with $\Omega$
taken real. The pulse duration $T$ is much shorter than the
lifetime of the atomic upper state, hence atomic damping  can be
discarded. Introducing the rotating frame operators

\begin{equation}
\hat{\sigma_{\pm}}(t)=\hat{S_{\pm}}(t)e^{\mp i\omega_l t}, ~\quad
\hat{\sigma_z}(t)\equiv\hat{S_z}(t),
\end{equation}
where  the $\hat{\sigma}$ operators  obey the same algebraic form
of Eq.(\ref{Com}), Heisenberg equations for the atomic operators
$\hat{\sigma}_{\pm,z}$  according to (\ref{Ham}) are of the form,
\begin{eqnarray}\label{motion}
\dot{\hat{\sigma}}_{+}&=
&i\Delta\hat{\sigma}_{+}-i\Omega(t)\hat{\sigma}_z(1+e^{-2i\omega_lt})
=\Big[\dot{\hat\sigma}_{-}\Bigl]^\dag,
\nonumber\\
\dot{\hat\sigma}_{z}&=&-i\frac{\Omega(t)}{2}
\Bigr[\hat{\sigma}_{+}(1+e^{2i\omega_l
t})-\hat{\sigma}_{-}(1+e^{-2i\omega_l t})\Bigl],
\end{eqnarray}
where $\Delta= \omega_a-\omega_l$, is the atomic detuning. Not
that the terms in $e^{\pm2i\omega_l t}$ represent the effect of
the interaction of the atom with the pulse outside RWA( note that,
the issue of RWA is only associated with linearly polarized light;
cf\cite{Shore}). \vspace{0.3cm}\\  {\bf(a)~Exact solution within
RWA.}

 Within RWA we discard the terms $e^{\pm2i\omega_l t}$ in
(\ref{motion}) and assume the qubit initially in the coherent
state,
\begin{equation}
\ket{\theta,\phi}=\cos(\frac{\theta}{2})\ket{0}+e^{-i\phi}\sin(\frac{\theta}{2})\ket{1},
\end{equation}
where $0\leq \phi\leq 2\pi$, $0\leq \theta \leq \pi$ and
$\ket{0},\ket{1}$ are the bare ground and  excited atomic states
respectively. Using the notation,
\begin{equation}
u_i(0)=\bra{\theta,\phi}\hat{\sigma_i}(0)\ket{\theta,\phi},~
i=x,y,z
\end{equation}
 for  the initial  Bloch vector where
$\hat\sigma_{\pm}=\hat\sigma_x\pm i\hat\sigma_y$, the exact
solution  of Eq.(4)in terms of the Bloch vector $u_i(t)$ within
RWA are written in the following form \cite{shukry},
\begin{equation}
\left(
\begin{array}{c}
u_x^{(0)}(t)\\
u_y^{(0)}(t)\\
 u_z^{(0)}(t)
\end{array}
\right)= \left(
\begin{array}{ccc}
\alpha^{(0)}_x&\alpha^{(0)}_y&\alpha^{(0)}_z \\
\beta^{(0)}_x&\beta^{(0)}_y&\beta^{(0)}_z\\
 \gamma^{(0)}_x&\gamma^{(0)}_y&\gamma^{(0)}_z
 \end{array}
 \right)
 \left(
 \begin{array}{c}
 u_x(0)\\
u_y(0)\\
 u_z(0)
 \end{array}
 \right),
\end{equation}
where
\begin{eqnarray}
\alpha^{(0)}_x&=&
\frac{1}{2}\Bigl[\left(\frac{\Omega}{\Omega_1}\right)^2+\left(\frac{\Delta^2+\Omega_1^2}{\Omega_1^2}\right)\cos\Omega_1
t+\left(\frac{\Omega}{\Omega_1}\right)^2(1-\cos\Omega_1 t)\Bigr],
\nonumber\\
\alpha^{(0)}_y&=&-\left(\frac{\Delta}{\Omega_1}\right)\sin\Omega_1
t,
\nonumber\\
\alpha^{(0)}_z&=&\left(\frac{\Delta\Omega}{\Omega_1^2}\right)(1-\cos\Omega_1
t),
 \nonumber\\
\beta^{(0)}_x&=&-\alpha^{(0)}_y,
\nonumber\\
\beta^{(0)}_y&=&
\frac{1}{2}\Bigl[\left(\frac{\Omega}{\Omega_1}\right)^2+\left(\frac{\Delta^2+\Omega_1^2}{\Omega_1^2}\right)\cos\Omega_1
t-\left(\frac{\Omega}{\Omega_1}\right)^2(1-\cos\Omega_1 t)\Bigr],
\nonumber\\
\beta^{(0)}_z&=&-\left(\frac{\Omega}{\Omega_1}\right)\sin\Omega_1
t,
\nonumber\\
\gamma^{(0)}_x&=&\alpha^{(0)}_z, \quad
\gamma^{(0)}_y=-\beta^{(0)}_z,
\nonumber\\
\gamma^{(0)}_z&=&\left(\frac{\Omega}{\Omega_1}\right)^2\Bigl[(\cos\Omega_1
t+\left(\frac{\Delta}{\Omega}\right)^2\Bigr],
\end{eqnarray}
with $\Omega_1=\sqrt{\Omega^2+\Delta^2}$  and we have used the
notations $u^{(o)}_{x,y,z}(t)$  are used for the exact RWA
solution.
 \vspace{0.3cm}\\
  {\bf (b)~Iterative solution outside
RWA.}

 Outside RWA, Eq.(\ref{motion}) has  an iterative solution
to $O(\lambda)$; $\lambda=\frac{\Omega}{\omega_o}$ at exact
resonance $(\Delta=0)$ and is given by \cite{shukry},
\begin{equation}\label{without}
\left(
\begin{array}{c}
u_x^{(1)}(t)\\
u_y^{(1)}(t)\\
 u_z^{(1)}(t)
\end{array}
\right)= \left(
\begin{array}{ccc}
\alpha^{(1)}_x&\alpha^{(1)}_y&\alpha^{(1)}_z \\
\beta^{(1)}_x&\beta^{(1)}_y&\beta^{(1)}_z\\
 \gamma^{(1)}_x&\gamma^{(1)}_y&\gamma^{(1)}_z
 \end{array}
 \right)
 \left(
 \begin{array}{c}
 u_x(0)\\
u_y(0)\\
 u_z(0)
 \end{array}
 \right),
\end{equation}
where,
\begin{eqnarray}
\alpha^{(1)}_x&=&1+\lambda\Bigl[(2-\frac{1}{4}\sin2\omega_l
t)\sin\Omega t+(1-\frac{1}{4}\cos2\omega_l t)\cos\Omega t\Bigr],
\nonumber\\
\alpha^{(1)}_y&=&-\frac{\lambda}{4}\Bigl[(1-\cos2 \omega_l
t)\sin\Omega t-\sin 2\omega_l t\cos\Omega t\Bigr],
\nonumber\\
\alpha^{(1)}_z&=&\frac{\lambda}{2}\left(\cos 2\omega_l t
\cos\omega t-1\right),
\nonumber\\
\beta^{(1)}_x&=&\frac{\lambda}{4}\Bigl[(1-\cos2 \omega_l
t)\sin\Omega t+\sin 2\omega_l t\cos\Omega t\Bigr],
\nonumber\\
\beta^{(1)}_y&=&\cos\Omega t-\frac{\lambda}{4}\Bigl[\sin 2\omega_l
t\sin\Omega t+(4-\cos2\omega_l t)\cos\Omega t\bigr],
\nonumber\\
\beta^{(1)}_z&=&-\sin\Omega t-\frac{\lambda}{2}\sin2\omega_l
t\cos\Omega t,
\nonumber\\
\gamma^{(1)}_x&=&\lambda(2-\cos2 \omega_l t-\cos\Omega t),
\nonumber\\
\gamma^{(1)}_y&=&\sin\Omega t+\lambda\sin2\omega_l t,
\nonumber\\
\gamma^{(1)}_z&=&\cos\Omega t.
\end{eqnarray}
and have  used the notations $u^{(1)}_{x,y,z}(t)$  for the
iterative solution outside RWA. Note, for $\lambda=0$, the
solutios $(9)$ with $(10)$ reduce to those in $(7$) with $(8)$ at
$\Delta=0$.

\section{Information transfer}

 Let us assume that we have coded
information in the initial state of the qubit. The initial density
operator takes the form,
\begin{equation}\label{init}
\rho(0)=\frac{1}{2}\Bigl(1+u_x(0)\hat{\sigma_x}+u_y(0)\hat{\sigma_y}+u_z(0)\hat{\sigma_z}\Bigr).
\end{equation}
 Due to the interaction with the pulse the   coded information may be
 exchanged between the pulse and the initial  state in (\ref{init}). Our
aim to is find how information evolves with time. We measure the
transfer information using the fidelity,
$\mathcal{F}=tr\{\rho(t)\rho(0)\}$ \cite{Ping}, while the entropy
exchange, $\mathcal{S}_e=-tr\{\rho ln\rho\}$ is used as a measure
of the information exchanged between the system and the
environment \cite{Xiang} . For this puropose we plot some
figures for different initial state settings within and outside RWA.\vspace{0.3cm}\\
{\bf  (a)Within  RWA}.

 In Fig.(1), we investigate the effect of the initial phase
angle $\phi$, on the fidelity of the transfer and exchange
information. Fig.(1a) shows the periodic  dynamical behavior  of
the fidelity in the resonant case i.e $\Delta=0$. The initial
state of the coded information is chosen such that
$\ket{\psi(0)}=e^{-i\phi}\ket{1}$. It is clear that, for
$\phi=\frac{\pi}{2}$ the fidelity, $\mathcal{F}$ decreases with
time smoothly  and completely vanishes for the first time round at
$\Omega t\simeq 2.5$. As one decreases $\phi$ the fidelity
decreases but does not completely vanish and the minimum values of
the fidelity $\mathcal{F}$,  decrease as $\phi$ decreases .

\begin{figure}
  \includegraphics[width=15pc,height=11pc]{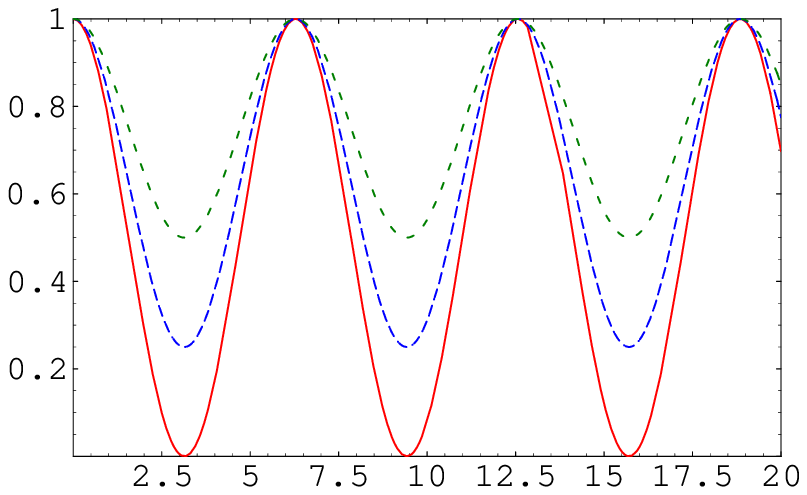}~\quad
   \includegraphics[width=15pc,height=11pc]{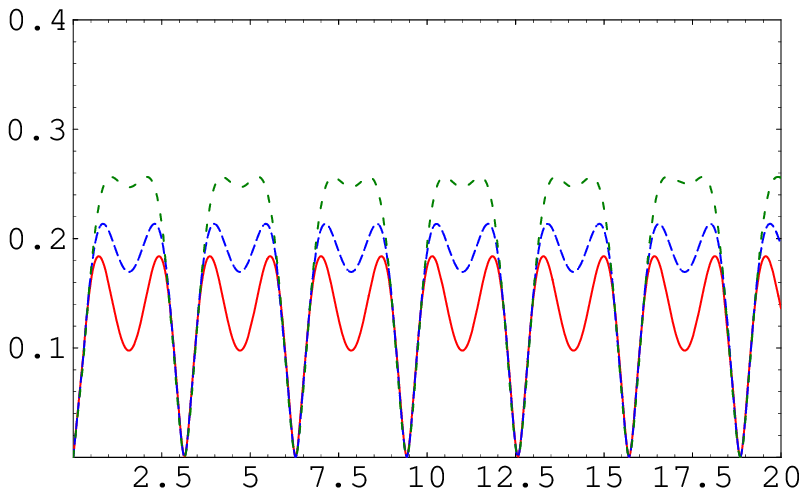}
    \put(-375,73){$\mathcal{F}$}
     \put(-190,75){$E_n$}
      \put(-350,118){(a)}
       \put(-160,118){(b)}
       \put(-80,-10){$\tau$}
       \put(-280,-10){$\tau$}
    \caption{The Fidelity $\mathcal{F}$  and the exchange information $E_n$, $(a),(b)$ respectively
   against the scaled time
  $\tau=\Omega t$ for the system
  (within RWA), $\lambda=0$ with $\Delta=0$ and $\theta=\frac{\pi}{2}$. The solid, dash and dot
  curves are
  for  $\phi=\frac{\pi}{2},
  \frac{\pi}{3}$ and $\frac{\pi}{4}$ respectively.}
\end{figure}

In Fig.(1b), we investigate the amount of information exchanged
between the state and the environment for the same parameters as
in Fig.(1a). As the fidelity decreases the exchange information
increases but does not reach  the maximum value i.e there is still
coherent information in the sate which carries these coded
information.  Also, with the fidelity $\mathcal{F}$ reaches to its
minimum value there is no exchange information. This means that,
for some interval of time one can use the driven qubit safely in
quantum communication.

\begin{figure}
  \begin{center}
  \includegraphics[width=15pc,height=11pc]{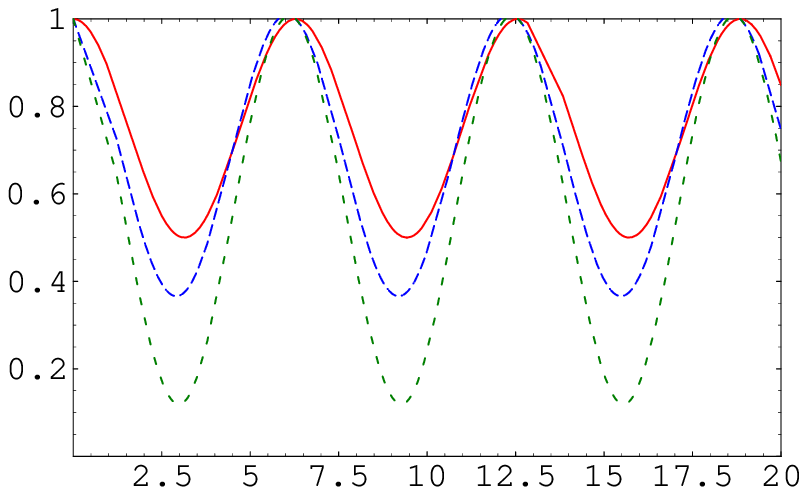}~\quad
   \includegraphics[width=15pc,height=11pc]{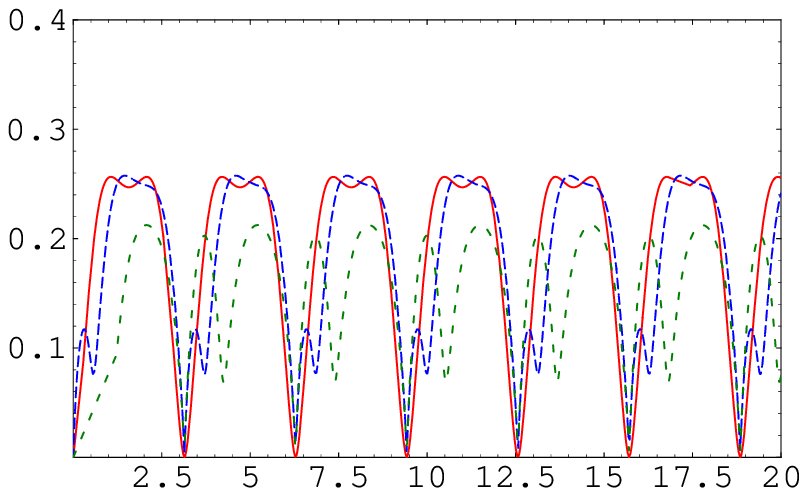}
    \put(-375,73){$\mathcal{F}$}
     \put(-190,75){$E_n$}
      \put(-350,118){(a)}
       \put(-160,118){(b)}
       \put(-80,-10){$\tau$}
       \put(-280,-10){$\tau$}
   \caption{The same as Fig.(1), but for fixed $\phi=\frac{\pi}{4}$ and different
   $\theta=\frac{\pi}{3},\frac{\pi}{4},\frac{\pi}{6}$ for the  solid, dashed and dot curves  respectively.}
       \end{center}
\end{figure}

In Fig.(2), we consider different  initial states setting for
$\theta$, where the phase angle is fixed at $\phi=\frac{\pi}{4}$.
In this case the initial state of the driven qubit is defined by
$\ket{\psi(0)}=\cos(\theta/2)\ket{0}+sin(\theta/2)e^{-i\pi/4}\ket{1}$.
 In general the
behavior of the fidelity is the same as that depicted in Fig.(1a),
but the the fidelity does  not completely vanish as in Fig.(1a)
for some classes of initial states. As $\theta$ decreases , the
minimum value of $\mathcal{F}$ decreases. Also, the exchange
information, Fig.(2b), increases  for large fidelity and
vice-versa. At the minimum values of the fidelity the exchange
almost vanishes.

\begin{figure}
  \begin{center}
  \includegraphics[width=15pc,height=11pc]{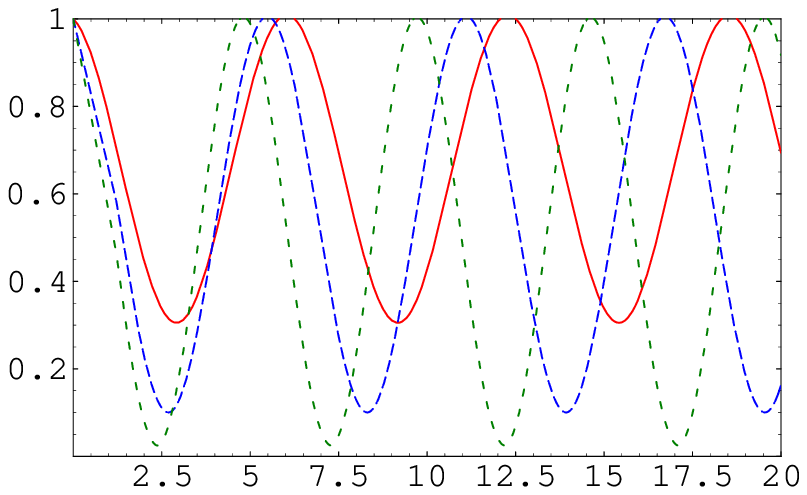}~\quad
   \includegraphics[width=15pc,height=11pc]{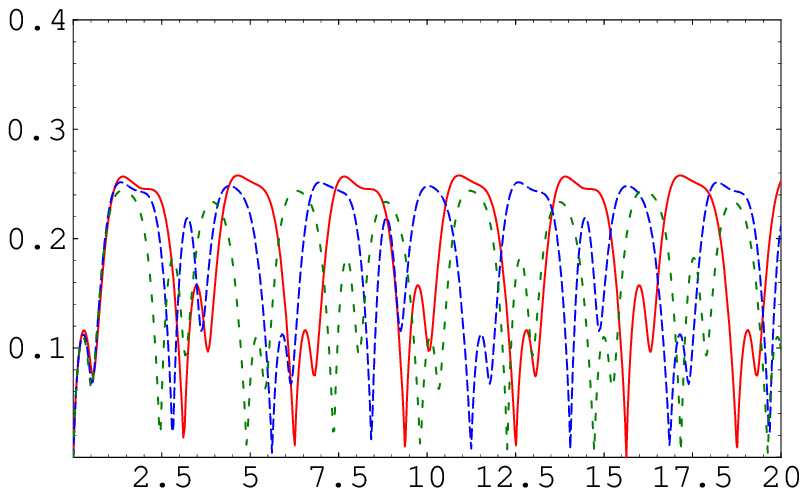}
     \put(-375,73){$\mathcal{F}$}
     \put(-190,75){$E_n$}
      \put(-350,118){(a)}
       \put(-160,118){(b)}
       \put(-80,-10){$\tau$}
       \put(-280,-10){$\tau$}
   \caption{The same as Fig.(1) but for
  $\theta=\frac{\pi}{3}$,$\phi=\frac{\pi}{4}$. The
  solid, dashed and dot curves  for $\Delta=0.1,0.5 $ and $0.8$  respectively.}
       \end{center}
\end{figure}

\begin{figure}
  \begin{center}
  \includegraphics[width=15pc,height=11pc]{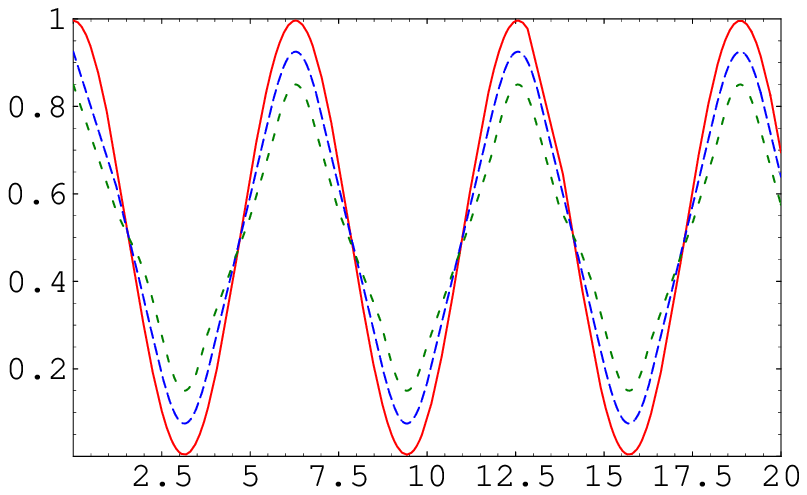}~\quad
   \includegraphics[width=15pc,height=11pc]{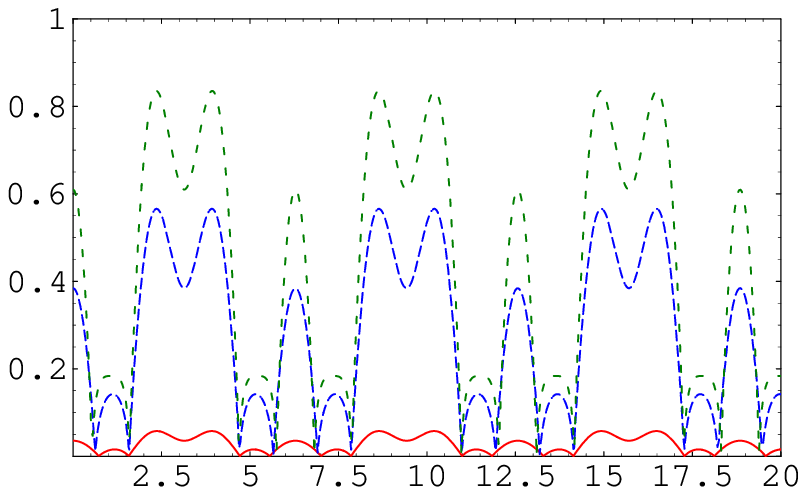}
    \put(-375,73){$\mathcal{F}$}
     \put(-190,75){$E_n$}
      \put(-350,118){(a)}
       \put(-160,118){(b)}
       \put(-80,-10){$\tau$}
       \put(-280,-10){$\tau$}
  \caption{(a)The fidelity $\mathcal{F}$ and (b) the information exchange $E_n$ against $\tau$ for
the  system treated outside the RWA ($\lambda\neq 0$). The initial
state of the qubit is described by
  $\ket{\theta=\phi=\frac{\pi}{2}}$ and $\Omega=\omega=1$. The  solid, dashed and dot curves  for $\lambda=0.01,0.2 $ and
$0.4$ respectively.}
       \end{center}
\end{figure}

The effect of the detuning parameter, $\Delta$ is  seen in
Fig.(3), where we consider the   choice for $\theta=\frac{\pi}{3}$
and $\phi=\frac{\pi}{4}$ which maximize the fidelity and minimize
the change information. In this case the initial state in which we
coded the information is given by
$\ket{\psi(0)}=\frac{\sqrt{3}}{2}\ket{0}+\frac{1}{2}e^{-i\frac{\pi}{4}}\ket{1}$.
The general behavior of the fidelity is the same as shown in
Fig.(1) and Fig.(2). For a very small value of the detuning
$\Delta=0.1$, the minimum  value of $\mathcal{F}$ is  lesser than
the resonance case: see the solid curve in Fig.(2a) and Fig.(3a).
As one increases the value of $\Delta$, the curves shift to the
left. This means that the fidelity decreases faster at earlier
time. Fig.(3b) describes the dynamics of the exchange information,
where  for small values of the detuning the exchange operation
starts  with delay. The maximum amount of information for
different values of the detuning  shows slight change.
\vspace{0.3cm}\\

{\bf (b) Outside the RWA.}

 The behavior of the fidelity
$\mathcal{F}$ and the information exchange ${E}_n$ for the system
given by (\ref{without}) where the system is treated outside RWA
is shown in Fig.(4). In this figure we set
$\theta=\phi=\frac{\pi}{2}$ (the case  described by the solid
curve in Fig.(1a), within RWA). The dynamics of the fidelity is
displayed in Fig.(4a) for different values of the parameter
$\lambda$. It is clear that for small value of $\lambda (=0.01)$,
the behavior  is almost the same as that within RWA. However, for
larger values of $\lambda$, the minimum value of the fidelity
increases. This means that the loss of the coded information
decreases. Fig.(4b) describes the dynamics of the exchange
information between the travelling qubit and the environment for
the same parameter as that used in Fig.(1a). This figure shows
that the amount of information exchange between the qubit and the
environment is much larger compared with that depicted in
Fig.(1b)( solid-curve within RWA). Also,the minimum value of $E_n$
increases for larger values of $\lambda$.
\begin{figure}[b!]
  \begin{center}
  \includegraphics[width=15pc,height=11pc]{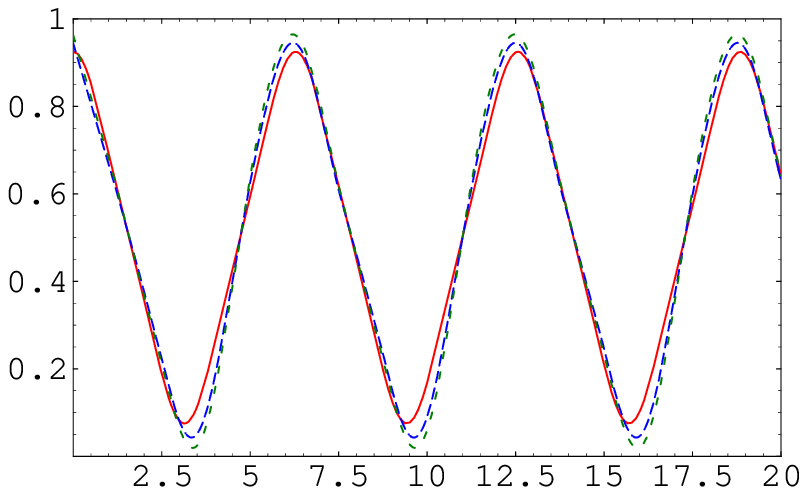}~\quad
    \includegraphics[width=15pc,height=11pc]{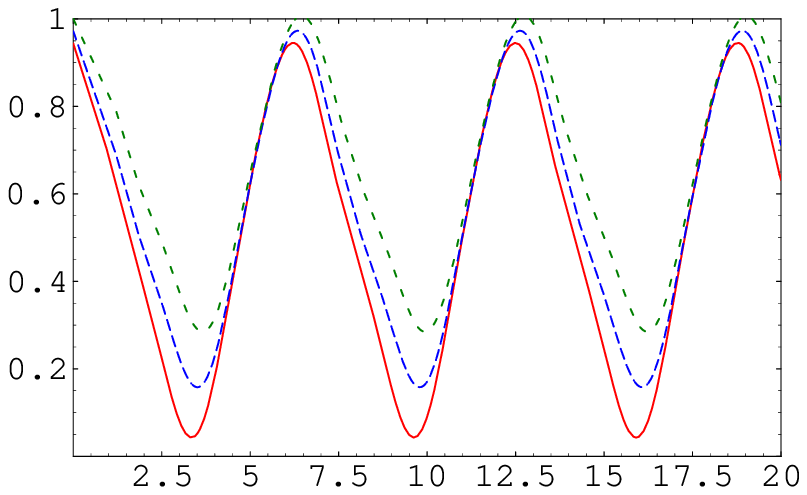}
    \put(-375,70){$\mathcal{F}$}
     \put(-190,70){$\mathcal{F}$}
      \put(-350,118){(a)}
       \put(-160,118){(b)}
       \put(-80,-10){$\tau$}
       \put(-280,-10){$\tau$}
  \caption{(a) The fidelity $\mathcal{F}$ against $\tau$ outside RWA $(\lambda=0.2)$ for
  fixed $\phi=\frac{\pi}{2}$ and different $\theta=\frac{\pi}{2},\frac{\pi}{3},\frac{\pi}{4}$
  (solid, dashed and dot curves  respectively)
  (b) The same as (a) but for fixed  $\theta=\frac{\pi}{2}$ and
  different  $\phi=\frac{\pi}{2},\frac{\pi}{3},\frac{\pi}{4}$( solid, dashed and dot
  curves respectively)}.
    \end{center}
\end{figure}
The effect of the excitation angle $\theta$ and the phase $\phi$
is investigated in Fig.(5), where we set the iteration parameter
$\lambda=0.2$. The dynamics of the fidelity $\mathcal{F}$ is shown
in Fig.(5a) for different values of the excitation angle $\theta$.
At $t=0$ the fidelity $\mathcal{F}<1$, due to the effect of the
$\lambda$- parameter $\lambda$. However, as one decreases the
value of $\theta$, the upper and lower values of the fidelity show
slight change. Fig.(5b), displays the behavior of $\mathcal{F}$
for different phases. As the phase decreases the minimum value of
the fidelity is much larger and consequently the exchange
information between the travelling qubit and the environment
decreases. So, by decreasing the value of the phase one can
overcome the negative effect of the iteration parameter $\lambda$.

Therefore, from our preceding results the fidelity of the
transmitted information, which is coded in an initial state
described by small excitation  angle $\theta$ and phase $\phi$
decreases faster but the maximum value of the exchange information
with the environment is smaller. However, the minimum value of
fidelity loss can be improved for larger values of the angle
$\theta$  and the phase $\phi$. If we consider a small value of
the detuning parameter, we can send information with high
fidelity, while the exchanged information is always upper bounded.
Outside the RWA  the fidelity of the travelling qubit could
improve, but at the the same time increases the possibility of
information exchange with the environment.  The effect of the
excitation angle $\theta$ outside RWA has a slight effect and the
fidelity doesn't reach its maximum value. However, for a smaller
phase angle $\phi$, the value of $\mathcal{F}=1$.

\section{Orthogonality Speed}

In the pervious section, we showed how the coded information
evolves from one locations (node) to another \cite{Mar}.
 In a quantum computer there it is  important to know  the speed of sending information from  one node to
 another.
Since the information is coded in a density operator, therefore we
seek how fast the density operator will change its orthogonality .
In other words, we search for a minimum time needed for a quantum
system to pass from one orthogonal state to another
\cite{Mar,Borrs}.

 Let us assume that the eigenvectors of the
initial state are given by $u_1(0)$, $u_2(0)$  and for the final
state are given by $u_1(t), u_2(t)$.
 The speed of quantum computation is defined by  the
maximum number of orthogonal states that the system can pass
through per unit time. The  orthognathy is given by the scalar
product of the vectors \cite{Yung,Fac}
\begin{equation}
Sp_{ij}=<u_i(0)|u_j(t)>, ~\quad i,j=1,2.
 \end{equation}
 \vspace{0.3cm}\\
 {\bf (a) Within  RWA.}

In Fig.(6), we have plotted $|Sp_{ij}|$, the amplitude values of
$Sp_{ij}$ against the scaled time to display its behavior for
different initial  atomic state settings.  From our results in
Sec.$3$, we know that for small values of $\theta$ and $\phi$, one
can transmit the coded information with high degree of fidelity.

In Fig.(6a), in   the resonant case $(\Delta=0)$ and
$\theta=\frac{\pi}{2}$ and a very small value of $\phi=\pi\times
10^{-3}$, it is clear that $|Sp_{ij}|$ vanishes periodically at
some specific time, $\Omega t$. At these times, the initial and
final states are orthogonal and the information which  is coded in
the initial state is completely transfered to the final state. As
one increases the phase angle $\phi=\frac{\pi}{10}$, the speed of
orthogonality decreases (Fig.(6b)), where the number of vanishing
$|Sp_{ij}|$ is lesser than that for smaller phase (see Fig.(6a)).
Fig.(6c), is plotted for different class of initial state
settings, where we consider $\theta=\frac{\pi}{4}$. It is seen
that the number of orthogonality decreases and consequently the
speed of computations is smaller than that depicted in Fig.(6a).
The non-resonant case is shown in Fig.(6d), for $\Delta=0.7$ where
 $|Sp_{ij}|$ does not vanish periodically as that predicted for
the resonant case (see Fig.(6a)). As an example, around $\omega
t\simeq 25 $ and $ 30$, $|Sp_{ij}|\neq 0$. This means that the
information has not completely transferred to the final state and
consequently there are some information gained by
the environment.\vspace{0.3cm}\\
{\bf (b) Outside RWA.}

\begin{figure}[t!]
  \begin{center}
  \includegraphics[width=15pc,height=11pc]{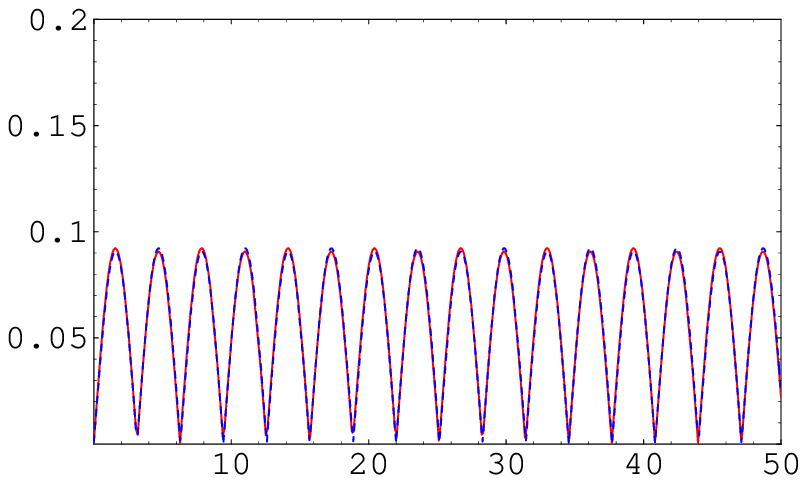}~\quad
    \includegraphics[width=15pc,height=11pc]{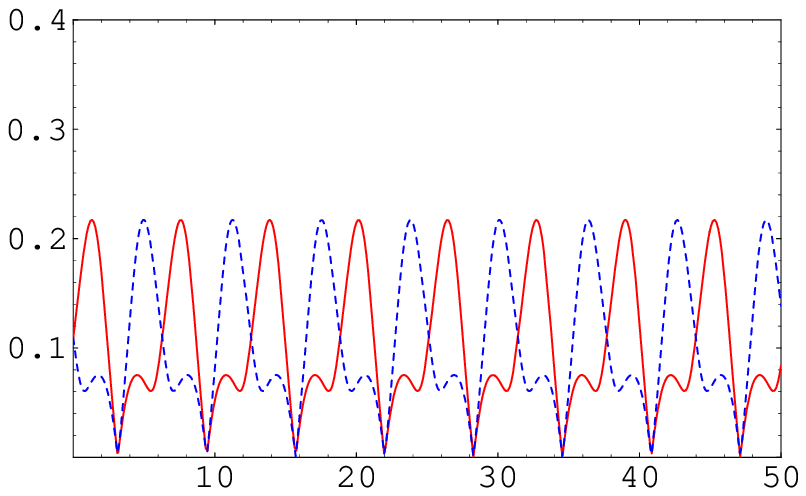}
     \put(-350,118){(a)}
       \put(-160,118){(b)}
        \put(-390,70){\small$|Sp_{ij}|$}
  \put(-195,70){\small$|Sp_{ij}|$}
  \put(-80,-5){$\tau$}
       \put(-280,-5){$\tau$}\\
\includegraphics[width=15pc,height=11pc]{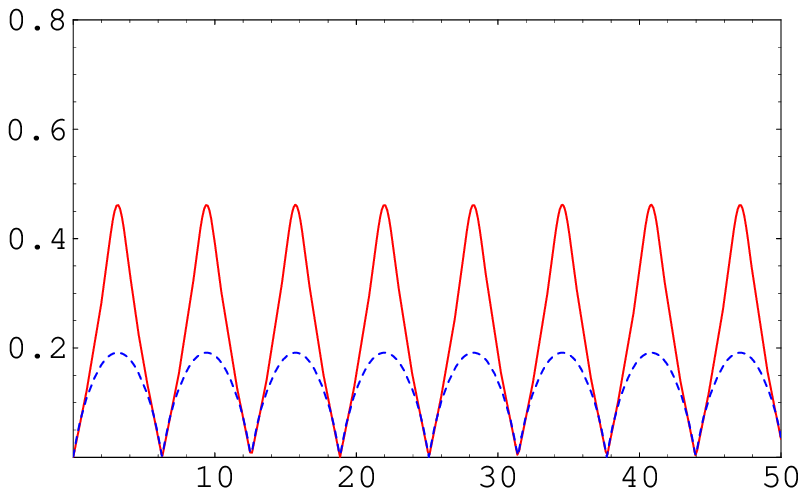}~
 \includegraphics[width=15pc,height=11pc]{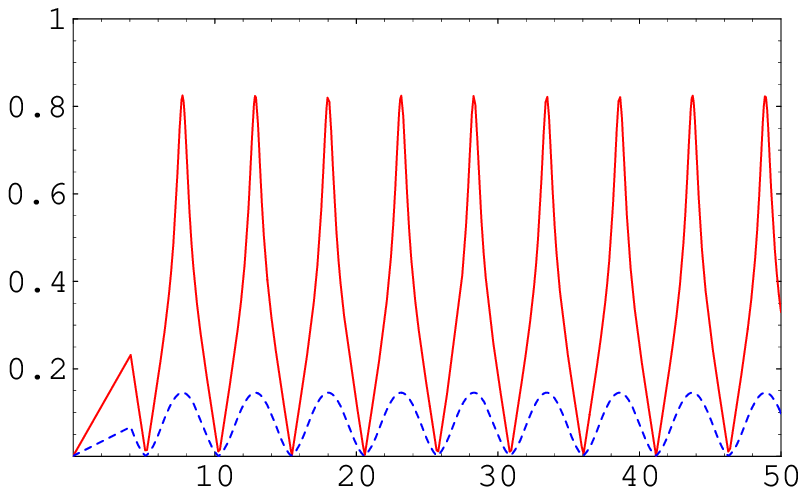}
                \put(-350,118){(c)}
       \put(-160,118){(d)}
       \put(-80,-5){$\tau$}
       \put(-280,-5){$\tau$}
       \put(-390,70){\small$|Sp_{ij}|$}
  \put(-195,70){\small$|Sp_{ij}|$}
  \caption{The speed of orthogonality of the qubit as a function of the scaled
  time $\tau$. (a) For $\theta=\frac{\pi}{2}, \phi=10^{-3}\pi$,
  $\Delta=0$. (b) The same as (a), but $\phi=10^{-1}\pi$. (c) the same as(a), but $\theta=
  \frac{\pi}{4}$. (d) The same as (a), but for $\Delta=0.7$.}
       \end{center}
\end{figure}

\begin{figure}[t!]
  \begin{center}
  \includegraphics[width=15pc,height=11pc]{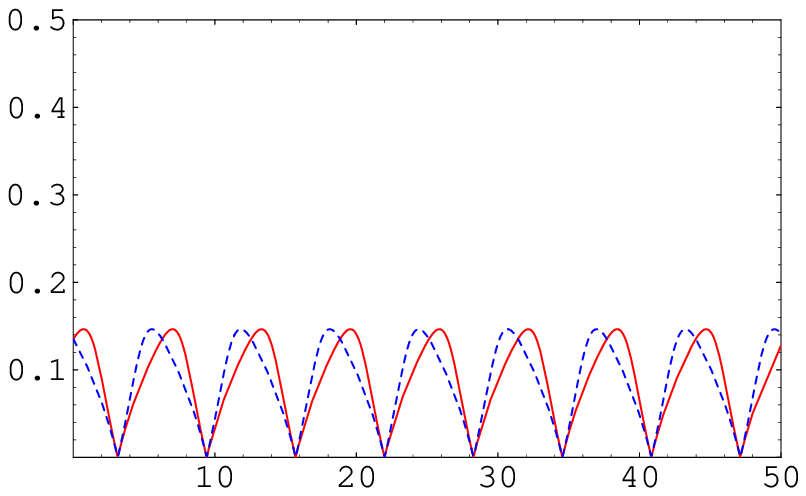}~\quad
    \includegraphics[width=15pc,height=11pc]{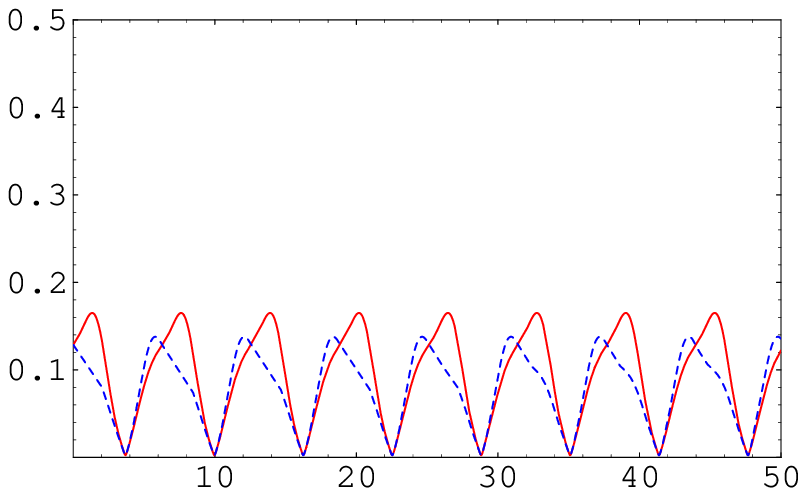}
     \put(-350,118){(a)}
       \put(-160,118){(b)}
        \put(-390,70){\small$|Sp_{ij}|$}
  \put(-195,70){\small$|Sp_{ij}|$}
  \put(-80,-5){$\tau$}
       \put(-280,-5){$\tau$}\\
\includegraphics[width=15pc,height=11pc]{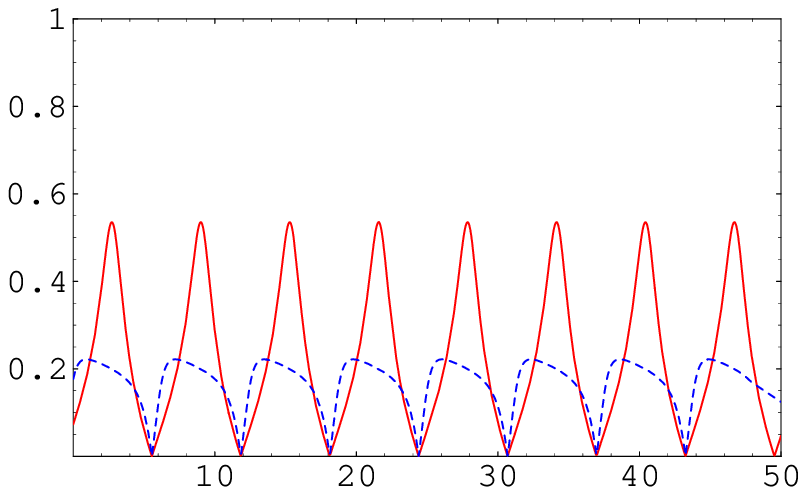}~\quad
 \includegraphics[width=15pc,height=11pc]{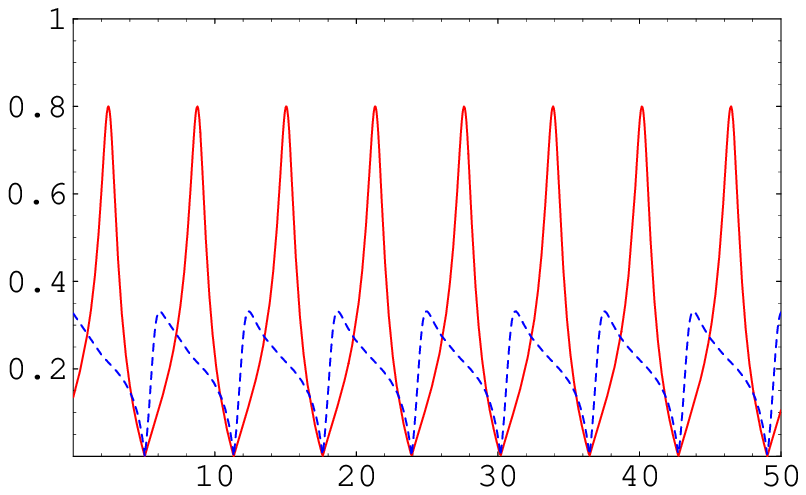}
                \put(-350,118){(c)}
       \put(-160,118){(d)}
       \put(-80,-5){$\tau$}
       \put(-280,-5){$\tau$}
       \put(-390,70){\small$|Sp_{ij}|$}
  \put(-195,70){\small$|Sp_{ij}|$}
 \caption{The orthogonality speed for  the qubit  treated outside the RWA ($\lambda\neq
  0$)vs
  time, $\tau$. (a) For $\theta=\frac{\pi}{2}, \phi=\frac{\pi}{8}$ and
  $\lambda=10^{-4}$.
(b) The same as Fig.(a), but $\lambda=0.08$. (c) The same
as(a),but $\theta=
  \frac{\pi}{4}$ and $\phi=\frac{\pi}{8}$.
  (d) The same as (a), but $\theta=\phi=\frac{\pi}{4}$.}
       \end{center}
\end{figure}

The dynamics of the orthogonality speed for a system treated
outside RWA is displayed in Fig.(7). The effect of the iteration
parameter $\lambda$ is shown in Figs.(7a$\&$7b), where we set
$\lambda=10^{-4}$ and $0.08$ respectively. It is clear that, the
orthogonality speed  $|Sp_{ij}|$ increases  for small value of
$\lambda$. As one decreases the excitation angle $\theta
(=\frac{\pi}{8})$, with  other parameters kept the same as in
Fig.(7a), the speed of orthogonality decreases (see Fig.(7c)).
However for large value of the phase angle, the orthogonality
speed decreases- compare Figs.(7a$\&7d$).

\section{Conclusion}
The dynamics of the coded information  in  a pulsed-driven  qubit
is investigated  within  and outside the rotating wave
approximation (RWA). Within RWA, the initial  atomic state setting
plays the central role for the fidelity of sending information
from one location to another. The fidelity and the exchange
information are  increased by decreasing the phase $\phi$ and the
polarized angle $\theta$ of the initial atomic coherent state
$\ket{\theta,\phi}$. The sensitivity of the fidelity of the
transmitted information to the detuning parameter is discussed,
where for large values of detuning, the fidelity decreases faster.
However the maximum value of the exchange information between the
qubit and the environment is slightly affected  as one increases
the detuning parameter. Also, we have investigated  the effect of
the initial state setting and the detuning  on the speed
orthogonality where it is shown that, the driven qubit could be
used to achieve quantum computations much faster.

Outside RWA, we have examined  the effect of the iteration
parameter $\lambda$ on the dynamics of the exchange information
and the fidelity of the transmitted information.  Although the
minimum value of the fidelity increases as one increases the value
of $\lambda$, the exchange information between the travelling
information and the environment increases. The excitation angle
($\theta)$ has a small effect on the dynamics of the fidelity of
the travelling information. However, for smaller value of the
phase ($\phi$)the minimum value of the travelling information's
fidelity increases and its maximum value reaches one.
 The speed of orthogonality, for the  system
treated outside RWA is investigated for different values of
$\lambda$, excitation and  phase angles $(\theta, \phi)$. We show
that, the orthogonality speed  decreases  for large values of
$\lambda$ and the initial phase $\phi$. However the speed of
orthogonality decreases as one decreases the excitation angle
($\theta$).

\end{document}